\documentstyle[prl,aps,epsf,epsfig,multicol]{revtex}    

\begin{document}

\bibliographystyle{plain}

\title{Disordered Heteropolymers with Crosslinks -  Phase
Diagram and Conformational Transitions}
\author{Lorin Gutman and Eugene Shakhnovich}
\address{ Department of Chemistry and
Chemical Biology, Harvard University, Cambridge, MA 02138}
\maketitle
\begin{abstract}
We study the phase behavior of random heteropolymers (RHPs) with
quenched cross-links, a novel polymer class of technological and
biological relevance, and show the possible occurrence of freezing
with few chain conformations sampled.  The sensitivity of the frozen
phase microstructure to the disorder components is elucidated at
positive solubility parameter values; at low T's segregated
microphases form, while at a finite T, a first order conformational
transition occurs, and is attributed to statistical matching of large
microphases bounded by cross-links.  The end of the symmetry broken
regime stabilization by cross-links occurs at a higher T by a second
order conformational transition.
\noindent ({\it submitted to Chem. Phys. Lett.})
\noindent 
\end{abstract}
\vspace{-.8cm}
\begin{multicols}{2}
Random Heteropolymers (RHP) with some monomers bonded by quenched
cross-links may be viewed as a natural offspring of two structurally
remote classes of polymers, cross-linked homopolymeric networks
\cite{kn:gn1}$-$\cite{kn:gn3}, and linear RHPs
\cite{kn:gn4}$-$\cite{kn:gn6}, both renown for their impact on
technology and biology.  The theoretical analysis of cross-linked RHPs
is eminently challenging; two frozen disorders are carried by the
polymer; the first disorder component is the sequence distribution of
disparate monomers that has been fixed during synthesis: for linear
RHPs known manifestations of having a fixed sequence of segments in
polymer properties are mesophase formation due to spatial grouping of
monomers with similar composition \cite{kn:gn4}$-$\cite{kn:gn5}, and
the reduction of overall chain conformations to a small number of
predominant folds (frozen phase formation)
\cite{kn:gn6}$-$\cite{kn:gn9}.  Analytical
\cite{kn:gn4}$-$\cite{kn:gn9} and computer simulations
\cite{kn:gn10}$-$\cite{kn:gn13} studies of linear RHP chains clarified
major questions on protein folding possible mechanisms, thermodynamics
and provided insight into the dynamic pathways of folding of proteins
to their native conformation.  The main past study emphasis on
crosslinked chains has been on crosslinked homopolymers, the earliest
proeminent work in the field has been by Deam and Edwards
\cite{kn:gn1}; recent advances attained in the understanding of
cross-linked homopolymer networks have been discussed and reviewed by
Goldbart and co-workers \cite{kn:gn3} and Panyukov \cite{kn:gn2}.

A detailed modeling of crosslinked RHPs is challenging, while 
complex phase behavior is expected in comparison with linear
RHPs; the quenched sequence distribution of segments fixed by the
chain connectivity of the incipiently linear RHP was shown
\cite{kn:gn6} to preclude macrophase separation by formation of
microdomains.  Crosslink formation in linear RHPs, and the nature of
these crosslinks (quenched or annealed) are
expected to significantly impact the formation and stability of
microphases, and their interdomain interfaces as well as the overall
number of possible accessible chain folds.

The weaker disorder - the sequence distribution of segments
reminiscent of linear RHPs, and the strong disorder component - the
fixed junction network reminiscent of crosslinked homopolymers, both
present in our system $\underline{cannot}$ be treated as annealed
disorders.  This important issue is addressed in the present work by
the extension of spin glass methodology of analysis of systems with
one quenched disorder \cite{kn:gn23}; our formalism allows the
explicit analytic treatment of the quenched nature of both the
sequence and crosslink distribution.  
The phase behavior, the
conformation chain organization in these phases, the interrelation of
spatial monomer separation by composition to the occurrence of phases
of few dominant folds, and the transition between these phases is
studied.

{\bf Theory Development}

Imagine a dilute solution of statistical two-letter
heteropolymers. Composition specific crosslinking agents are
introduced in solution and intrachain composition specific junctions
form; the A and B segments residing in proximal spatial range can form
homogeneous crosslinks (A-A and B-B type) and heterogeneous ones (A-B
type). Upon completion of the crosslink synthesis, the crosslinks are
quenched by chemical means and no further crosslink reorganization can
occur. For example in proteins cross-links are quenched by irreversible reactions
with iodoacetamide or by acidification \cite{kn:gn28}.

The microscopic interactions for the incipient two-letter linear RHP are
described by a continuous microscopic Hamiltonian representation
inspired from Edwards work on homopolymers \cite{kn:gn18}. {\bf r}(n)
is the spatial location of the $n$'th segment, while $\theta(n)$ monitors
the chemical composition of the n'th segment.  $\theta(n)=1$ for an A
segment and $\theta(n)=-1$ for a B segment.  The coarse-grained
description of the segment composition fluctuations for statistical
RHPs along the chain contour obeys a Gaussian process with mean,
$<\theta>=2f-1$, and sequence fluctuations
\cite{kn:gn20}$-$\cite{kn:gn21},
$<(\delta\theta(n)\delta\theta(n'))>$=$\delta(n-n')4f(1-f)l$; l is the
statistical segment length, and f is the fraction of A segments.
The Hamiltonian is:
\begin{eqnarray}
 H'_{RHP}=\frac{3}{2l}\int\dot{\bf r}^{2}dn 
+\frac{1}{2}\sum_{i,j=A,B}\int 
d{\bf r}\int d{\bf r'}
\hat{\rho}_{i}({\bf r}){\bf V}_{ij} \hat{\rho}_{j}({\bf r})
\label{eq:lagrange}
\end{eqnarray}
The first term in eq. \ref{eq:lagrange} represents the nearest
neighbor interaction and accounts for the chain flexibility. 
In the second term of eq. \ref{eq:lagrange}, $\hat{\rho}_{i}({\bf r})$
represents the microscopic density composition components given by:
\begin{eqnarray}
\hat{\rho}_{A}({\bf r}) = \int dn \frac{1}{2}\left(1 + \theta (n)\right)\delta ({\bf r} - {\bf r}(n))\hspace{.1in};\nonumber\\
\hat{\rho}_{B} ({\bf r}) = \int dn \frac{1}{2}\left(1 - \theta (n)\right)\delta ({\bf r} - {\bf r}(n)) \label{eq:vfields}
\end{eqnarray}
${\bf V}_{ij}$, the binary inter-segment interactions, are represented
by the following matrix:
\begin{eqnarray}
\overline{\bf V}_{ij}= \left( \begin{array}{cc} V_{AA}({\bf r-r'}) & V_{AB}({\bf r-r'})\\
V_{BA}({\bf r-r'}) & V_{BB}({\bf r-r'}) \end{array} \right)
\label{eq:Vintmatrix}
\end{eqnarray}
$V_{AA}$({\bf r-r'}), $V_{AB}$({\bf r-r'}) and $V_{BB}$({\bf r-r'})
represent A-A, A-B, and B-B segment-segment interactions,
respectively. 
Quenched composition specific crosslinks of type A-A, B-B, or A-B are described
in our theory as instantaneous spatial constraints imposed on the partition function of a
linear RHP.  
A general composition specific crosslink has the form
$\delta({\bf r}_{i}(n)-{\bf r}_{j}(n'))$ where i,j =A,B.

Fluctuations in the total number of cross-links between different
disorder realizations of crosslinks are allowed by a Poisson distribution; this choice has been recently discussed
$\cite{kn:gn16}, \cite{kn:gn17}, \cite{kn:gn29}$ in context of homopolymer network
studies.  Thus, a fixed number of A-A, A-B
and B-B cross-links is linearly parameterized by $\mu_{AA}, \mu_{BB}$
and $\mu_{AB}$ , and the probability of fluctuations around $\mu_{AA}, \mu_{BB}$
and $\mu_{AB}$ is given by:
\begin{eqnarray}
P[\mu's]=\frac{[\mu_{AA}]^{M}}{M!}\frac{[\mu_{AB}]^{J}}{J!}\frac{[\mu_{BB}]^{K}}{K!}\nonumber\\exp[-(\mu_{AA}+\mu_{BB}+\mu_{AB})]\label{eq:poisson}
\end{eqnarray} 
Thus, the partition function of an RHP chain with a fixed sequence and
constrained by M crosslinks of type A-A, K crosslinks of type B-B, and
J cross-links of type A-B is given by:
\begin{eqnarray}
Z[\theta(n)]= \underline{\int}\overline{\int} D{\bf r}(n)
exp(- \frac{3}{2l}\int\dot{\bf r}(n)^{2}dn+ \nonumber\\ 
\frac{1}{2}\sum_{i,j=A,B}\int 
d{\bf r}\int d{\bf r'}
\hat{\rho}_{i}({\bf r}){\bf V}_{ij}({\bf r-r'}) \hat{\rho}_{j}({\bf r}))\nonumber\\
\left[\frac{1}{2^{M}M!} (\int d{\bf r}(\hat{\rho}_{A}({\bf r})^{2})^{M}\right]
\left[\frac{1}{2^{K}K!}(\int d {\bf r} \hat{\rho}_{B}({\bf r})^{2})^{K}\right] 
\nonumber\\ \left[\frac{1}{J!} (\int d{\bf r}\hat{\rho}_{A}({\bf r})\hat{\rho}_{B}({\bf r}))^{J}
\right]\label{eq:partit}
\end{eqnarray}
In eq. \ref{eq:partit} we have expressed the crosslinking constraints by microscopic
sequence-dependent-composition-densities
(viz. eq. \ref{eq:vfields}). The spatial {\bf r} integration and the
coefficients in front of the constrains accounts for formation of all
possible crosslinking constraints that are consistent with
crosslinking from an equilibrium ensemble of composition specific
contacts, and reminiscent of the typical experimental chemical crosslinking
process considered here. 

The free energy computation of our problem requires to average the
logarithm of the partition function, Z given in eq. \ref{eq:partit},
since both the sequence and the crosslinks are quenched:
\begin{eqnarray}
F= \sum_{[\theta],[r_{i}]}P_{1}([\theta])P_{2}([\theta],[r_{i}])log(Z([\theta], [r_{i}]))
\end{eqnarray}
$[\theta]$ represents one fixed sequence realization while $[r_{i}]$
are the spatial coordinates of one fixed crosslink realization.
$P_{1}([\theta])$ is the probability distribution for the synthesis of
one quenched sequence, while $P_{2}([\theta],[r_{i}])$ is the
conditional probability distribution of one specific realization of
composition specific cross-links given a preexisting fixed sequence;
thus at each given sequence realization, crosslinks can form by fixing
some composition specific intersegment contacts from spontaneously
occuring chain conformations adopted by the linear RHPs.

The crosslink average is performed first by introducing 3(n+1)
identical copies of the system; this mathematical trick originally
introduced by Deam and Edwards \cite{kn:gn1} in their studies of
crosslinked homopolymers, is an exact and non-perturbative way for
fixing intersegment contacts from spontaneously occuring binary
contacts from an equilibrium distribution of chain conformations in
homopolymers \cite{kn:gn1}:
\begin{eqnarray}
F([\theta])= \sum_{[r_{i}]}P_{2}([r_{i}],[\theta])log(Z([\theta],
[r_{i}])= \nonumber\\\left(\frac{\partial}{\partial n}\right)_{n
\rightarrow{ 0}}log <(Z^{n+1})>_{cl}\label{eq:av}
\end{eqnarray}

Our definition of this averaging trick is slightly different from Deam
and Edwards formulation; in our formulation the derivative of log(Z)
allows in the present problem an exact representation of the sequence
dependent denominator occuring due to the probability normalization,
in a nominator form, and sets the ground for an exact analytical
sequence average later on.

The crosslink average of the log of the partition function given in
eq. \ref{eq:partit} is performed with the Poisson distribution given
in eq. \ref{eq:poisson} by summation over M, J, L values.

Next we perform the average over disordered sequence; the log
generated in the crosslink average (viz. eq. \ref{eq:av}) is averaged
now over the sequence by using one more time our averaging trick;
thus, the free energy is now given by:
\begin{eqnarray}
F= \left(\frac{1}{m}\right)_{m \rightarrow
0}\left(\frac{\partial}{\partial n}\right)_{n \rightarrow 0}
<<(Z^{(n+1)\otimes m})>>_{sq, cl}\label{eq:double}
\end{eqnarray}
Interestingly, the sequence average of the log generates in turn a new index m of copies.

All together, $m \otimes (n+1)=l$; $l$ is the total number of copies
of our system.  Methods developed in the field of spin glasses by
Parisi \cite{kn:gn24} are extended herein, and used to obtain an
analytical solution.  The sequence distribution average is carried out here exactly
in the usual way \cite{kn:gn20}.

As in previous RHP studies, cross-linked RHPs in compact states are of
interest; under these conditions a scaling argument shows \cite{kn:gnJCP} that
it is reasonable to compute the free
energy by a one-step mean-field Parisi \cite{kn:gn24} calculation. It
was shown that the reduction in the total number of chain
conformations to a small number of dominant folds, a phenomena usually
termed in the RHP literature as freezing into a few chain
conformations, is described by one order parameter $\xi$; the polymer
coordinates, (the annealed degrees of freedom) equilibrate in response
to the quenched disorder realizations regardless of the disorder
source, sequence, crosslinks or both; thus, in the present case one freezing order
parameter $\xi_{0}$ measures the total reduction of chain conformations due to
energetic and entropic constraints. The physical meaning of $\xi$ is as follows: 
\begin{eqnarray}
\xi_{0}=1-\sum p_{i}^{2}\label{eq:parametrization}
\end{eqnarray}
$p_{i}$ is the probability of the i'th chain conformation. 
For many chain conformations, each
conformation has a low probability of occurrence, and $\xi_{0}$ practically equals
one. If the chain is collapsed in one conformation,
 $p_{i}$=1, while for other chain conformations where i$\neq$ j,
$p_{j}$=0 implying that $\xi_{0}$=0.

In the present case, not like in the linear RHP case, $\xi$
is an explicit function of the indices, $(\alpha,k)$.  A suitable one
step parameterization of the freezing order parameter $\xi$, is
$\xi_{0}= x_{0} x_{0}'$; m and n form a tensor space, $l=m\otimes n$;
$x_{0}$ and $x_{0}'$ are the one step continuous parameterizations
\cite{kn:gn24} of the indices $\alpha$ (crosslinks sector), and k
(sequence sector), respectively.  A separate publication is now in
print wherein all mathematical details and derivations are presented
\cite{kn:gnJCP}.  The scenario of $x_{0}'=1$, $0<x_{0}<1$ is defined
as sequence-induced-freezing, the scenario $x_{0}=1$,
$0<x_{0}'<1$ is defined as crosslink-induced-freezing while the
parameter regime of $0<x_{0}<1$, $0<x_{0}'<1$ is defined as
sq.+cl.-induced-freezing as obviously inferred by the parameterization
relevant in determination of the overall number of folds.  These
convenient definitions allows to keep a good track of the numerical
results while we discuss the physics but should not be taken too
literally; first, since the true order parameter that monitors the
occurrence of few dominant folds is $\xi_{0}$ while $x_{0}$ and
$x_{0}'$ are parameterizations only; also, as our calculations show,
and is inferred by the crosslinking procedure the disorder
components are strongly correlated, and it is not possible to
completely separate the component disorder manifestations in the conformational
organization of the crosslinked RHP.

Using Parisi commutation relations \cite{kn:gn24}, the total free
energy per monomer for an RHP in compact state with composition
specific and quenched cross-links is:
\begin{eqnarray}
F= \frac{\gamma}{x_{0}x_{0}'}(1-2(x_{0}+x_{0}') + \nonumber\\
(2x_{0}'-1)(x_{0}-1)(x_{0}-2)^{2})\nonumber\\ + \frac{log(1-\mu_{a}
x_{0}x_{0}')}{x_{0} x_{0}'} -{\cal V}(\mu 's, f)(x_{0}-1) \nonumber\\
\mu_{a}=\frac{\sigma^{2}}{2}\rho\overline{\chi
(\mu's)}\hspace{.01in};\hspace{.01in}\gamma=\frac{l}{0.25(2V_{AB}+V_{AA}+V_{BB})}
\label{eq:free_energy1}
\end{eqnarray}
with:
\begin{eqnarray}
{\cal V}(\mu's)=0.5(\mu_{AA}+\mu_{BB}+2\mu_{AB})\nonumber\\
{\cal V}(f,\mu's)=\nonumber\\0.5{\cal V}(\mu's)-0.25(\mu_{AA}+\mu_{BB})
\overline{\theta}-0.25\chi_{Fcl}(\mu's)\overline{\theta}^{2}\nonumber\\
\chi_{F}=0.5(2V_{AB}-V_{AA}-V_{BB})\nonumber\\ 
\chi_{Fcl}(\mu's)=0.5(2\mu_{AB}-\mu_{AA}-\mu_{BB})\nonumber\\
;\overline{\chi(\mu's)}=\chi_{F} - \chi_{Fcl}(\mu's)\nonumber\\
\label{eq:interactions}
\end{eqnarray}
Note the important consistency requirement that our parameterization obeys:
for $x_{0}'$=1., the free energy in
eq. \ref{eq:free_energy1} is tantamount the free energy of an RHP with
annealed crosslinks, a problem that has been recently analyzed
\cite{kn:gn27}.  Next, we compute numerically the stability of F with
respect to $x_{0}$ and $x_{0}'$ in all parameter regimes. While the
value of $\xi_{0}$ obtained is used to determine the occurrence of few
dominant folds, the stability of the free energy and the value of the
parameterizations of $\xi_{0}$, $x_{0}$ and $x_{0}$', provides
essential information on the sensitivity of the microdomain structure
of the frozen phase to the quenched sequence distribution and the
fixed crosslink realizations. Let us now explicitly explore the
conformational organization of RHPs in the globular phase.  The
scenario depicted in fig. 1 corresponds to having a small number of
cross-links of type A-B, and a positive $\chi_{f}$ (segments are
encouraged to group with alike). The physics displayed by fig. 1 with
regard to the conformational organization of the RHPs is qualitatively
illustrated in fig. 2.  At low T's, few dominant RHP folds occur due
to formation of energetically driven microphases with segregated
domain structure and having sharp interfaces.  The heterogeneous
cross-links formation, following the crosslinking procedure, nucleates at the A-B
interfaces of the segregated microphases, and reduces 
the microphase interfacial free energy since each crosslink is an 
entropic constraint; thus chain reorganizations in this regime
is most sensitive to the crosslink disorder component.
The spatial microphase organization within the low temperature frozen
phase is qualitatively represented in fig. 2; a heterogeneous
cross-link is marked by two adjacent small circles.  Thus at low
temperatures, freezing occurs in the crosslink sector of the free
energy; we term this regime as a crosslink-frozen-globular phase.  Our
numerical calculations of the free energy stability shows that at
$T_{1}$ (fig. 1), the chemical potentials for the formation of
crosslink-frozen-globular and sequence-frozen-globular phases become
identical.
\begin{figure}[ht]
\vspace{1.cm}
\hspace{-2.5cm} \psfig{file=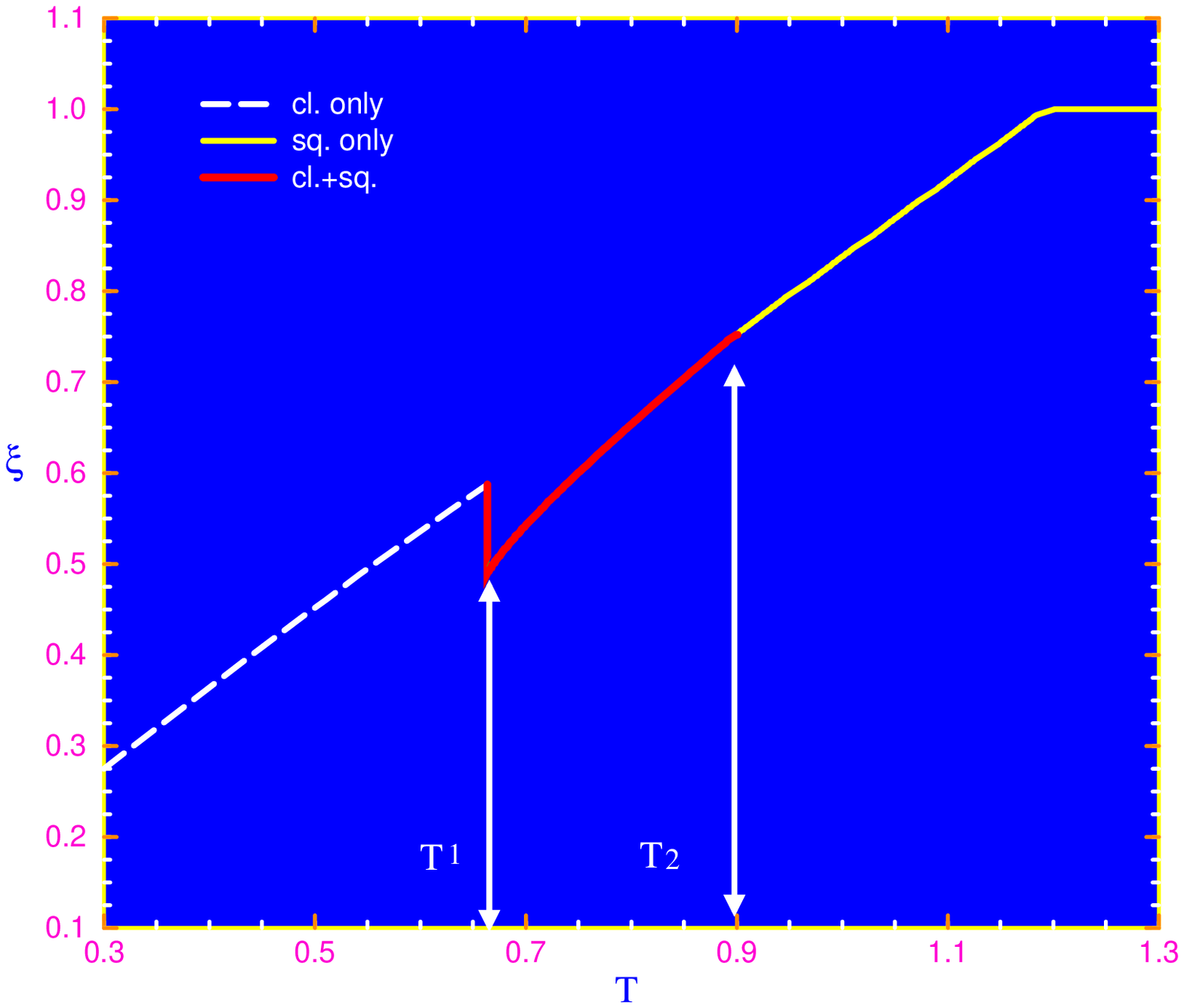,width=6.cm,angle=270}
\vglue-0.8cm
\vspace{0.1cm}
{\small Fig. 1: Frozen phase analysis of RHPs with
composition specific and quenched crosslinks; Variation of the
freezing order parameter $\xi_{0}$ with temperature. Results for l=1,
$\mu_{A-A}$=0, $\mu_{B-B}$=0, $\mu_{A-B}$=0.01, $\chi_{F}$=2.,
$\rho$=1., $v_{0}$=0.2 and f=0.5}
\label{cv70}
\end{figure}
\vspace{-0.6cm}
\begin{figure}
\epsfysize=2.5in \epsfbox{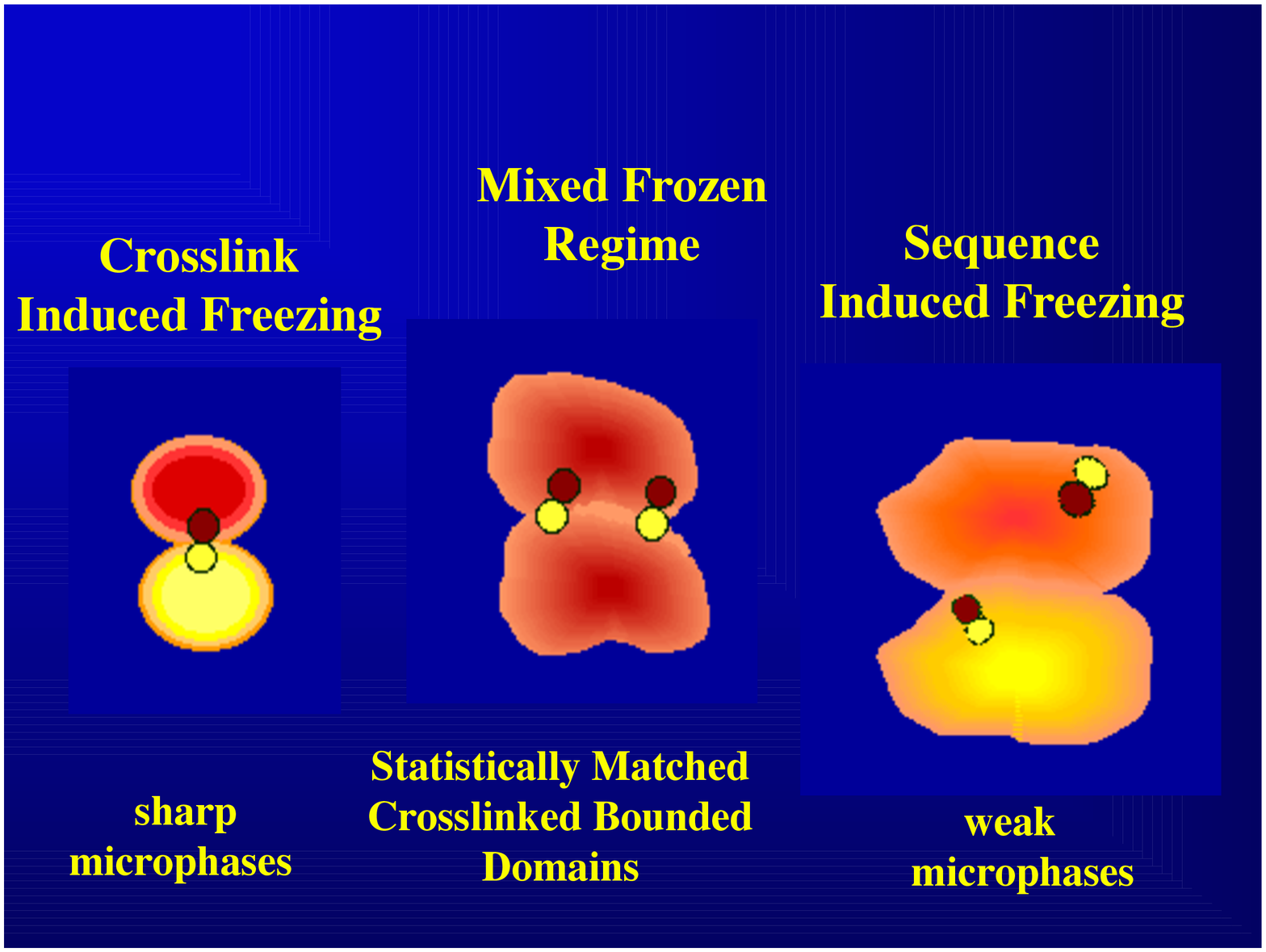}
\vglue0.01cm
\small{Fig. 2: Qualitative illustration of crosslink location among
segregation patterns emerging within the frozen phase. Red and yellow
regions depict microphases of A and B segments respectively, while the
color intensity denotes the segregation strength.  
Two small adjacent circles represent two crosslinked monomers;
red small circle represents a monomer of type A, while
a yellow small circle represents a monomer of type B.} 
\end{figure}

In the language of critical phenomena this would imply the
occurrence of a coexistence regions between two frozen-globular RHP
phases with an equal number of folds in each phase dominated in one
phase by the sequence disorder and in the other phase by the
cross-links disorder component. In the present system, $\xi_{0}$ is
the only physically meaningful order parameter that measures the
reduction in chain conformations, and thus monitors the occurrence of
freezing; $x_{0}$ and $x_{0}$' parameterize $\xi_{0}$. As a result,
the coexistence of separate individual such phases is not possible.
Our numerical calculation confirms this fact, and the ``coexistence''
point does not occur; at $T_{1}$, a new first order conformational
transition occurs to a more stable phase wherein the overall number of
few dominant folds is abruptly reduced, as marked by the spike in
$\xi_{0}$ (viz. red line in fig. 1); our numerical calculations show
that this new globular phase is characterized by the occurrence of
symmetry breaking induced by both crosslinks and sequence, thus we
name it as mixed-frozen-globular phase.  Let us now provide a physical
interpretation of this conformational transition. Below $T_{1}$, the
conformational organization of the RHP is primarily determined by the
inter-segment interaction strength. In the vicinity of $T_{1}$ the
existence of some cross-links induces formation of large domains
bounded by cross-links.  At the onset of formation of the
mixed-frozen-globular state at
$T_{1}$, the microphases are large in size and have diffuse interfaces
(viz. fig. 2 - the seq. + cl. induced freezing). Since the crosslinks
are heterogeneous, most likely they will form at this diffuse
interfaces, decreasing the interface flexibility and entropy. But the
penalty due to local deformation of large cross-linked bounded domains
is small, and a statistical pattern matching of microphases belonging
to separate cross-linked bounded domains occurs. The frozen globule
occurs here due to both sequence and crosslink disorder components;
only a few chain conformations allow prefered pattern matching of
microphases over length scales as large as the size of cross-linked
bounded domains, a realization which may explain the sharpness of the
spike in fig. 1.  The cooperativity observed here reminds to a large
extent of the cooperativity observed in the folding of disulfide
bonded proteins to their native state \cite{kn:gn28}.  In the folding
scenario of crosslinked proteins, crosslinks are also composition
specific but homogeneous, they nucleate within the segregated domains.
The apparently negative heat capacity inferred in fig. 2 is not
inconsistent with thermodynamics; fig. 1 displays the temperature
dependence of different thermodynamic systems and not phase behavior
variation within the same thermodynamic system.  This fact is a result
of the experimental crosslinking procedure.  At each temperature
within equilibrated globular phases of linear RHPs, quenched
crosslinks are formed. Thus at each T the system constraints (here the
quenched crosslinks) are different which implies that the phase
behavior comparison in fig. 1 is indeed between different
thermodynamic systems at different temperatures.  The complementing
future study scenario,  is the phase behavior of a linear RHP system
crosslinked at one temperature, and subject to temperature variations while the
crosslink realization of the initial crosslinking temperature is
retained.

Let us now return to the phase behavior analysis of fig. 1.  At
$T_{2}$ (viz. fig. 1), our numerical calculation shows that the
mixed-frozen-globule becomes unstable, while the
sequence-frozen-globular phase is stable and it should be observed as
it has the lowest chemical potential (viz. fig. 1 green line). The
occurrence of this continuous conformational transition is expected;
the number of cross-links is small while the reduction in the number
of dominant folds due to cross-links only cannot occur at high
temperatures. The $T > T_{2}$ frozen regime should be characterized by
weaker and diffuse microphases (viz. fig. 2 - sequence induced
freezing).  A good candidate for testing our predictions is the
ensemble growth Monte Carlo method \cite{kn:gn13}. This approach has
been successfully implemented in the study of linear RHPs in charged
disorder \cite{kn:gn14}, \cite{kn:gn15} and in confined geometries,
and allowed a faithful comparison with prior analytical calculations
on RHPs \cite{kn:gn2}, \cite{kn:gn6}.

{\bf Acknowledgments}

This work has been supported by NIH grant 52126.

\end{multicols}
\end{document}